\newcommand{\beqn}{\begin{eqnarray}}
\newcommand{\eeqn}{\end{eqnarray}}

\documentclass[
    ,final            
  ]
  {aipproc}

\layoutstyle{8x11double}

\begin{document}

\title{ Two-component quark-gluon plasma in stringy models }

\classification{11.25.Tq; 11.25.Wx; 12.38.Mh}
\keywords{quark-gluon plasma, holographic models}
\author{H. Verschelde}{address={\it $^\dag$ Ghent University, Department of Physics and Astronomy\\
Krijgslaan, 281-S9, 9000 Gent, Belgium
}}
\author{V.I. Zakharov}{
  address={ ITEP, B. Cheremushkinskaya 25, Moscow, 117218 Russia}
  ,altaddress={Max-Planck Institut f\"ur Physik, 80805 Muenchen, Germany
}}

\begin{abstract}
The two-component liquid model reproduces the
 basic properties of the quark-gluon plasma as observed in heavy-ion collisions.
 The key dynamic element of the model is the existence of a light scalar.
 We argue that existence of  such a scalar is a generic feature of
 stringy models of quantum chromodynamics. The lattice data provide
 evidence for a condensed, three-dimensional scalar field as well. We outline a possible
 crucial check of the model on the lattice.
\end{abstract}
\maketitle

Observations on the quark-gluon plasma
 at RHIC
\footnote{For details, discussions and references
 see, e.g.. reviews \cite{teaney}.}
 have led to the discovery of a quantum relativistic
 liquid at temperatures higher than the deconfinement phase
 transition, $T>T_c$. A crucial observation is the importance of quantum effects, as
 follows \cite{son1} from the low value of the ratio of the viscosity to the
 entropy density,
 \begin{equation}\label{low}
 {\eta\over s}~\approx~{1\over 4\pi}~,
 \end{equation}
 for an analysis of the data see \cite{romatschke}.
 There exist not many models of quantum liquids,
 and superfluidity is a natural first candidate.
 And, indeed, the two-component model, with one component being superfluid,
 explains `naturally' the basic observations on the plasma   \cite{chernodub}.

 The phenomenology of a two-component
liquid has been thoroughly discussed in the literature,
see in particular \cite{son2,yarom}.
 In the hydrodynamic approximation and neglecting dissipation effects
 one has the following basic expressions
 for the current and energy-momentum tensor:
\begin{eqnarray}\label{model}
j^{\mu}~=~nu^{\mu}+f^2\partial^{\mu}\phi\\ \nonumber
T^{\mu\nu}~=~(\epsilon+P)u^{\mu}u^{\nu}+P\eta^{\mu\nu}+
f^2\partial^{\mu}\phi\partial^{\nu}\phi\\ \nonumber
u^{\mu}\partial_{\mu}\phi~=~\mu \nonumber
\end{eqnarray}
where we used the notations of Ref \cite{yarom}.
Here,
$u^{\mu}$ is the 4 velocity of an element of the liquid,
representing the normal component,
$\phi$ is a {\it scalar} field, representing the superfluid component,
and $\mu$ is the chemical potential.

It is worth emphasizing, however,  that
the two-component picture has never been applied, to our knowledge,
directly to the actual data on the quark-gluon plasma. The resulting low
value of the viscosity,
see (\ref{low}), is then, in a way,
in contradiction with the classical (not quantum) nature of the
hydrodynamic model used. It would be very interesting to
check, whether inclusion of terms
containing derivatives from the scalar field  into $j_{\mu}$ and $T_{\mu\nu}$
changes significantly fits to the data.

In this paper we consider the possibility
that a variation of the famous two-component model
of superfluidity applies directly to the quark-gluon plasma
 The crucial issue is whether the QCD dynamics might produce a light scalar field
 entering (\ref{model}). The answer we get is rather in the affirmative. The stringy models
 of QCD tend to predict light scalars. A crucial phenomenological test of the model is mentioned.

\section{Scalar condensate}

\subsection{General constraints}

Dynamically, the validity of the superfluidity scenario depends strongly on the existence
of an (effective) scalar field $\phi$.
The only known way to keep a (real) scalar field massless is to assume
condensation of a complex field $\varphi$,
condensed into the thermal vacuum
\begin{equation}\label{condition}
\langle \varphi \rangle_{ground~state}~\neq~0~.
\end{equation}
The phase  $\phi$ of the field $\varphi$ corresponds then to a new light degree of freedom.

The condition (\ref{condition}) looks very restrictive and, in more detail, assumes a number
of constraints:

{\it a)} The field $\varphi$ is a complex field:
$$\varphi^{*}~\neq~\varphi~~.$$

{\it b)} There should be then a charge which
distinguishes the field $\varphi$. The thermal-vacuum expectation value
 (\ref{condition})  breaks spontaneously the corresponding symmetry.
  The problem is that in QCD we seemingly do not  have
   any symmetry of this type.

{\it c)} In the case of superfluidity, one thinks rather in terms of
 a   {\it three-dimensional} field $\varphi {\bf (r)}$ while the time derivative of
the corresponding phase
is determined by the chemical potential $\mu$ :
$\label{nonrelativistic}
\partial_t\phi~=~\mu~.
$
A relativistic generalization of this condition is included into (\ref{model}).

\subsection{Thermal scalar}

At first sight the conditions a)-c) above amount to a kind of no-go theorem for
superfluidity. However, it is striking that a 3d field with similar
properties arises naturally \cite{kogan} within a string model  and is commonly called the
thermal scalar. For a concise review and further insights see \cite{kruczenski} .

One considers temperatures $T$ below and close to the temperature of the
(Hagedorn) phase
transition $T_H$
\footnote{The temperatures of the Hagedorn and deconfinement phase transitions
coincide only for critical dimensions, $d=26$ for the bosonic strings ,
see \cite{kogan}. For us, however, only generic features of the
thermal scalar are important. We do not rely literally on existence of the
 thermal scalar, see below.}. In the string picture $\beta_H\equiv 1/T_H=4\pi\alpha'^{1/2}$
(where $1/(2\pi \alpha')\sim 1/l_s^2$ is the string tension). At $T=T_H$
the statistical sum over the states diverges. The main observation is that
at small $|T-T_H|$ the sum is dominated by the contribution of a single degree of freedom,
that is a complex scalar meson with the mass
\begin{equation}\label{mass}
m_{\beta}^2~\approx~{\beta_H(\beta_H-\beta)\over 2\pi^2(\alpha^{'})^2}~~,
\end{equation}
In other words, at $T=T_H$ the mass would become tachyonic.
There exist various, dual interpretations of the thermal scalar.
One way to visualize it is that (\ref{mass}) refers
to the mass of the mode once wrapped around the compact, Euclidean time
direction.

To consider the plasma we should address temperatures above the phase transition,
$T>T_H$ where (\ref{mass}) does not apply.
Imagine, however, that the thermal scalar becomes tachyonic and condenses at $T>T_H$.
Then, remarkably enough, the conditions we formulated above are
satisfied. Indeed:

a) The thermal scalar is a complex field.  This is because the string can be wrapped
around the Euclidean time coordinate in both directions, a typical $U(1)$ situation.

b) Thus, the thermal scalar is associated with the topological quantum number
which is the wrapping number around the compactified time direction. This is a quantum number
{\it specific} for strings. And this is a remarkable resolution of the puzzle
that we do need a symmetry to be spontaneously broken, on one hand, and
we cannot identify such a symmetry in the field-theoretic language on the other hand.
The dual-model language does allow for such an identification!

c) The thermal scalar is a 3d scalar field in the sense that
the time derivative of its phase is fixed:
\begin{equation}
|\partial_{\tau}\phi_{thermal~scalar}|~=~2\pi T~,
\end{equation}
Note an important change compared to the third equation
in (\ref{model}).

\subsection{Three dimensional scalar at $T>T_c$}

Nowadays, it is common to consider dual models of Yang-Mills theories
in terms of strings living in extra dimensions with non-trivial
geometry. The thermal scalar at temperatures below and close to $T_c$
is generic to such models as well,
see \cite{kruczenski}
and references therein.  However,  the phase transition is
treated now as a change of geometry
in the extra dimensions and the information on the scalar at $T<T_c$
does not directly help
to approach physics at  $T>T_c$. This is 'bad news'. 'Good news'
is that the 3d scalars are resurrected at $T>T_c$ in another disguise.

Very briefly, for details and
classical references see, e.g., \cite{gorsky}, there are
two compact directions, the Euclidean time and an extra one associated with
the $\theta$ dependence, or topological charge. As always, there is also
the scale-related
$z$ direction, with a horizon, $0<z<z_H$. At low temperatures the radius of
the time direction is independent of $z$ while the radius of the $\theta$-coordinate
tends to zero at the horizon:
\begin{equation}\label{geometry1}
R_{\tau}(z)~=~const~,~~R_{\theta}(z_H)~=~0~~;T<T_c~.
\end{equation}
At temperatures above the phase transition the geometry of the two
compact coordinates is interchanged so that:
\begin{equation}\label{geometry2}
R_{\tau}(z_H)~=~0~,~~R_{\theta}(z)~=~const~~;T>T_c~.
\end{equation}

As a result, one predicts that the defects become time-oriented
at $T>T_c$ \cite{gorsky}. In particular, the magnetic strings, well studied on the lattice at $T<T_c$ \cite{gorsky2},
are becoming time-oriented. The phenomenon can be readily understood from
the example of the thermal scalar. Indeed, the corresponding stringy mode corresponds
to wrapping around the time direction which fixes the time dependence
and, as a result, the thermal scalar is a $3d$ field \cite{kogan}.

The intersection of time-oriented strings with the spatial $3d$ volume is a set of
trajectories which are predicted to percolate in $3d$. In field theoretic
language, the $3d$ trajectories correspond to a $3d$ scalar field,
 let us call it magnetic field $\varphi_{magn}$. The percolation,
 in turn, corresponds to a non-vanishing vacuum expectation value:
\begin{equation}\label{condensate}
<\varphi_{magn}>^2~\sim~\Lambda_{QCD}~,
\end{equation}
which is a prerequisite for superfluidity of the gluon plasma (see above).
 The  independent  lattice results do support
 the validity of the prediction (\ref{condensate}),
 for detailed  analysis see \cite{nakamura}.

 \section{Possible crucial test of the model}

 So far, we listed arguments in favor of the two-component model of
 the plasma. However, one cannot claim, of course, that the data validate
 the model.
 A crucial test of the model could be performed through lattice measurements
 of a correlator of components of the energy-momentum tensor $T^{ti},i=1,2,3$.
 In more detail, consider the retarded Green's function defined as:
 \begin{equation}
 G_R^{tj,ti}(k)~\equiv~i\int d^4xe^{-ikx}\theta(t)\langle[T^{tj}(x),T^{ti}(0)]\rangle~~.
 \end{equation}
 Moreover, concentrate on the case of vanishing frequency, $k_0=0$.
 There are two independent form factors, corresponding to transverse and longitudinal waves.
 \begin{equation}
 G_R^{tj,ti}(0, {\bf k})~=~{k^ik^j\over {\bf k}^2}G^{L}_R({\bf k})+\Big(\delta^{ij}-
 {k^ik^j\over {\bf k}^2}\Big)G_R^T({\bf k})
 \end{equation}
In the absence of the superfluidity $G_R^T=G_R^L$ which means that there is no
non-analyticity at ${\bf k}\to 0$.

 Contribution of the superfluid component to the $G_R^{L,T}$ has been discussed in many
 papers and textbooks, see, in particular, Ref.  \cite{yarom}.
 which includes also relativistic corrections. The result is
 \begin{equation}
 \Big(\lim_{{\bf k}\to 0}[{G_R^T({\bf k})-G_R^L({\bf k})]}\Big)_{superfluidity}
 ~=~\rho_s\mu~.
 \end{equation}
where $\rho_s$ is the density of the superfluid component,
$\mu$ is the chemical potential.

In case of the gluonic plasma which we are considering a similar result
is expected to hold. An educated guess is:
\begin{equation}\label{test}
 \Big(\lim_{{\bf k}\to 0}[{G_R^T({\bf k})-G_R^L({\bf k})]}\Big)_{plasma}
 ~\sim~T (2\pi T)^2<\varphi_{magn}>^2~,
 \end{equation}
 where $T$ is temperature and $<\varphi_{magn}>^2$ is the vacuum expectation value
 of the magnetic scalar field discussed above.
 Again, a non-analytical term at ${\bf k}\to 0$ is predicted.

Note that the proposed crucial test of the model (\ref{test}) refers to static quantities
. Since there is no time dependence, the continuation from the Euclidean
to Minkowski space is straightforward and the prediction of the model, (\ref{test}),
can be tested on the lattice.

\section{Conclusions}

In more general terms, the stringy approach reveals mechanisms of generating
dynamical (i.e. not seen in the QCD Lagrangian) U(1) symmetries. They are
directly related to the topology of extra dimensions and   specific for strings.
Another lesson concerns applications of the holographic models.
According to conventional wisdom, the probability to find a defect
is exponentially suppressed in the limit of large $N_c$:
\begin{equation}
W_{defect}~\sim~\exp(-S_{defect})~,~~~S_{defect}~\sim~N_c~~.
\end{equation}
However, in the confining models it is generic that some radii
of extra dimensions are vanishing, see (\ref{geometry1}), (\ref{geometry2}).
Then, in the classical approximation there are defects whose action is vanishing:
$$S_{defect}~\sim~N_c\cdot 0~~,$$
for examples see \cite{gorsky}. The effect of such defects should be added
'by hand' to the standard formalism.

\begin{theacknowledgments}
We are thankful to Maxim Chernodub for collaboration at earlier stages of the
work on  this paper and discussions.
\end{theacknowledgments}

\bibliographystyle{aipproc}   

\begin{thebibliography}{99}
\bibitem{teaney}
 E. V. Shuryak,  \emph{Nucl. Phys.}   \textbf{750}  64 (2005),
[arXiv:hep-ph/040506];
 Th. Schafer, and  D. Teaney,
\emph{Rept. Prog. Phys. }  \textbf{72}   126001 (2009),
 arXiv:0904.3107 [hep-ph];
 D. Teaney, \emph{Prog. Part. Nucl. Phys.} \textbf{62}  451 (2009).

\bibitem{son1}
 P. Kovtun, D.T. Son, A.O. Starinets,
\emph{Phys. Rev. Lett.} \textbf{94} 111601 (2005),
[arXiv:hep-th/0405231].

\bibitem{romatschke}
P. Romatschke, and U. Romatschke,  \emph{Phys. Rev. Lett. } \textbf{99}  172301
(2007),  arXiv:0706.1522 [nucl-th].

\bibitem{chernodub}
M. N.  Chernodub, H.  Verschelde, V.I.  Zakharov,
arXiv:0905.2520 [hep-ph].



\bibitem{son2}
D.T. Son, {\it Int. J. Mod. Phys. A.} {\bf 16S1C} (2001), arXiv:hep-ph/0011246;
C.P. Herzog,   P.K. Kovtun,   D.T. Son,  \emph{Phys. Rev.} {\bf D79} 066002 (2009),
 arXiv:0809.4870 [hep-th].

\bibitem{yarom}
Ch.P. Herzog, and A.Yarom,  {\it Phys. Rev.} \textbf{D80} (2009) 106002,
arXiv:0906.4810 [hep-th].



 \bibitem{kogan}
J. J. Atick, and E. Witten, \emph{   Nucl. Phys.} \textbf{B310} 291 (1988);
B. Sathiapalan,  \emph{ Phys. Rev.} \textbf{ D35}  3277 (1988);
Ya. I. Kogan,  \emph{ JETP Lett.} \textbf{ 45}  709 (1987).

\bibitem{kruczenski}
M. Kruczenski, and  A. Lawrence,  \emph{ JHEP}  \textbf{0607} 031 (2006), [arXiv:hep-th/0508148].


\bibitem{gorsky}
A. S. Gorsky, V. I. Zakharov, A. R. Zhitnitsky, \emph{ Phys. Rev.} \textbf{ D79}
 106003 (2009),
 arXiv:0902.1842 [hep-ph].

\bibitem{gorsky2}
A. Gorsky,   V. Zakharov,  \emph{Phys. Rev.} \textbf{D77} 045017 (2008),
 arXiv:0707.1284 [hep-th].

 \bibitem{nakamura}
M.N. Chernodub, A. Nakamura, V.I. Zakharov,  arXiv:0904.0946 [hep-ph].


\end{thebibliography}


\IfFileExists{\jobname.bbl}{}
 {\typeout{}
  \typeout{******************************************}
  \typeout{** Please run "bibtex \jobname" to optain}
  \typeout{** the bibliography and then re-run LaTeX}
  \typeout{** twice to fix the references!}
  \typeout{******************************************}
  \typeout{}
 }

\end{document}